\documentclass{article}
\usepackage{spconf,amsmath,graphicx,hyperref}
\usepackage{times}
\usepackage{enumitem}
\usepackage{makecell}
\usepackage{booktabs}
\usepackage{xcolor}
\usepackage{fancyhdr}

\newcommand{\ieeecopyright}{%
  \footnotesize \textcopyright~2026 IEEE. Personal use of this material is permitted. Permission from IEEE must be obtained for all other uses, in any current or future media, including reprinting/republishing this material for advertising or promotional purposes, creating new collective works, for resale or redistribution to servers or lists, or reuse of any copyrighted component of this work in other works. Published in: DOI: 10.1109/ICASSP55912.2026.11461508
}
\fancypagestyle{firstpage}{%
  \fancyhf{}

  \fancyfoot[L]{\parbox{\textwidth}{\ieeecopyright}}
}

\title{CONTENT ANONYMIZATION FOR PRIVACY IN LONG-FORM AUDIO}

\makeatletter
\newcommand{\nlarge}{\@setfontsize\nlarge{11}{13}} 
\makeatother

\name{\nlarge Cristina Aggazzotti$^{\star}$ \qquad Ashi Garg$^{\star}$ \qquad Zexin Cai$^{\dagger}$ \quad Nicholas Andrews$^{\star \dagger}$\vspace{-1em}} 
  \address{\normalsize Johns Hopkins University, $^{\star}$Human Language Technology Center of Excellence,\vspace{-.2em}\\ \normalsize$^{\dagger}$Department of Computer Science, Baltimore, MD, USA \vspace{-.2em}\\ \normalsize\{caggazz1, agarg22, zcai21, noa\}@jhu.edu} 
      
\begin{document}
\ninept 
\maketitle
\thispagestyle{firstpage}

\begin{abstract}
\noindent Voice anonymization techniques have been found to successfully obscure a speaker's acoustic identity in short, isolated utterances in benchmarks such as the VoicePrivacy Challenge. In practice, however, utterances seldom occur in isolation: long-form audio is commonplace in domains such as interviews, phone calls, and meetings. In these cases, many utterances from the same speaker are available, which pose a significantly greater privacy risk: given multiple utterances from the same speaker, an attacker could exploit an individual's vocabulary, syntax, and turns of phrase to re-identify them, even when their voice is completely disguised. To address this risk, we propose a new approach that performs a contextual rewriting of the transcripts in an ASR-TTS pipeline to eliminate speaker-specific style while preserving meaning. We present results in a long-form telephone conversation setting demonstrating the effectiveness of a content-based attack on voice-anonymized speech. Then we show how the proposed content-based anonymization methods can mitigate this risk while preserving speech utility. Overall, we find that paraphrasing is an effective defense against content-based attacks and recommend that stakeholders adopt this step to ensure anonymity in long-form audio.
\end{abstract}

\section{Introduction}
Voice anonymization aims to mitigate privacy risks by modifying speech to conceal speaker identity while preserving utility. Existing approaches, driven by benchmarks like the VoicePrivacy Challenge, have shown success at the utterance-level~\cite{tomashenko2024voiceprivacy}. Simple methods such as kNN-based voice conversion can achieve strong anonymity even in the case of semi-informed attackers~\cite{li2024best}. However, these controlled, utterance-level benchmarks focus on \textit{utterance-level attacks}, while in practice a single audio sample may contain many utterances from the same speaker. Indeed, many real-world settings such as interviews, phone calls, and meetings typically contain many utterances from the same speaker~\cite{meyer2025use}.

The transition to long-form audio introduces a critical vulnerability: the linguistic content becomes a powerful biometric side-channel~\cite{aggazzotti2024can,aggazzotti2025asr}. Existing anonymization systems that focus exclusively on the acoustic signal or simple PII redaction leave the linguistic content untouched, since individual utterances do not contain enough linguistic information to be discriminative. This paper shows that the content renders current anonymization techniques insufficient for long-form audio and presents a novel method to address this fundamental challenge.

Our approach to joint content and voice anonymization is straightforward. Specifically, we propose to exploit the intermediate transcriptions of an automatic speech recognition (ASR) to text-to-speech (TTS) pipeline by rewriting the transcripts to eliminate speaker-specific style while preserving content. Paraphrasing is typically construed as a sentence-level task. However, performing utterance-by-utterance paraphrasing is challenging in the case of speech transcripts, since individual utterances may be quite short. This introduces several challenges. First, automatic paraphrasing systems may struggle to identify meaningful paraphrases of very short utterances (e.g., a single word). Second, the utterance structure itself may be revealing of the speaker, so a strong anonymization system may need to modify that structure, for example by removing backchannels or merging two short utterances into one. Finally, the intended meaning of an utterance may be uncertain without the broader context of the conversation.

\vspace{3pt} \noindent \textbf{Summary of contributions}  To address these challenges, we propose a contextualized paraphrasing model that operates on a sliding window of multiple utterances and rewrites them jointly, given the preceding context. While paraphrasing multiple utterances jointly is more computationally challenging, we show that small open-weight language models (\texttt{Gemma-3-4B}~\cite{team2025gemma}) can achieve close to the performance of API-based models like GPT-5. We address the utility of the generated speech, showing that content-based detection is no more effective than synthetic speech detection, and that the naturalness of the synthesized anonymized content is preserved. To our knowledge, we are the first to systematically evaluate LLM-based paraphrasing as a defense against content-based attacks in a voice anonymization pipeline, quantifying the privacy-utility trade-offs with modern generative models and detection systems.

\section{Related Work}

\noindent\textbf{Voice Anonymization} 
Voice anonymization aims to suppress speaker identity while retaining other useful attributes of speech. This provides a balance between privacy and utility, unlike encryption, which renders data unusable for downstream tasks, or redaction, which destroys utility. Research in this area is largely driven by benchmarks like the VoicePrivacy Challenge, which focus on anonymizing short, decontextualized utterances~\cite{tomashenko2024voiceprivacy}. While successful in this constrained setting~\cite{li2024best}, these methods overlook the challenges posed by long-form audio, particularly the threat from the linguistic content itself. 

\vspace{3pt}\noindent\textbf{Content Privacy} Some prior work has recognized the need for anonymization beyond voice~\cite{williams2021revisiting,williams2023new}, but thus far, subsequent work has focused on Personally Identifiable Information (PII), such as names and locations, usually masking~\cite{williams2024} or replacing ~\cite{turan2022,hui2025securespeech} any sensitive words or phrases. While these methods can remove overt identifiers, they do not address the more subtle stylistic indicators that can still be used to re-identify a speaker. Recently, work on \emph{written} text has shown that LLM-based rewriting can be an effective method for anonymizing writing style (i.e., authorship anonymization/obfuscation)~\cite{baocarpuat2024,fisher2024jamdec}; thus, we apply LLM-based paraphrasing to speech transcripts, also finding it useful for speaker obfuscation.  

\vspace{3pt}\noindent\textbf{Authorship Attribution} Most previous content-based attribution has focused on written language, often looking at frequencies of various stylometric features~\cite{neal2017,stamatatos2009} or embedding representations of style~\cite{rivera-soto2021,wegmann2022}. Features like vocabulary choice, syntax, and function word usage have been successfully used to re-identify individuals from text. Recent work pioneered the application of these techniques to automatically-transcribed transcripts, demonstrating that speaker identity can be recovered from speech content alone, even with transcription errors~\cite{aggazzotti2024can,aggazzotti2025asr}. This research confirms that the linguistic content of speech is a powerful biometric identifier that must be addressed by any robust anonymization system. Our proposed approach directly tackles this by using LLM-based paraphrasing to modify revealing stylistic features of a speaker.

\section{Preliminaries}\label{sec:prelim}

\noindent \textbf{ASR-TTS Anonymization Pipeline}
A common and effective voice anonymization method is a cascaded pipeline that combines automatic speech recognition (ASR) and text-to-speech (TTS) synthesis~\cite{hui2025securespeech}. In this approach, the original audio is first transcribed into text by an ASR system. This intermediate text representation provides a powerful opportunity to not only remove the original speaker's acoustic voiceprint but also modify the linguistic content. After any desired text-level modifications, a TTS system synthesizes new speech from the modified transcript using a different target-speaker voice. This process effectively decouples the original voice from the content.

\vspace{3pt}\noindent \textbf{Problem Statement}
The goal of voice anonymization is to transform a recording to conceal the original speaker's identity while preserving its utility. Formally, let $X = (u_1, u_2, \ldots, u_N)$ be a long-form audio recording from a source speaker, $s$. The defender's goal is to produce an anonymized version $X' = g(X)$ that is not attributable to $s$. The attacker has a scoring function, $f(X', E_s)$, to compute the similarity between the anonymized recording $X'$ and enrollment data $E_s$ for a known speaker. Successful anonymization is achieved when the attacker is maximally uncertain about the true speaker's identity, corresponding to an equal error rate (EER) of 50\%~\cite{tomashenko2024voiceprivacy, li2025attack}. Crucially, in the context of long-form audio, this function $f$ can be a \emph{multi-modal} model that leverages both acoustic evidence from the audio in $X'$ and linguistic evidence from its transcription. Therefore, successful anonymization must defend against both the acoustic and linguistic channels simultaneously. We focus on the \emph{uninformed} attacker setting in which the attacker does not have access to anonymized trials. 

\section{Methodology}
To address this multi-modal threat, our methodology centers on \emph{joint anonymization} of both the acoustic voice and the linguistic content. We use an ASR-TTS pipeline and introduce novel paraphrasing techniques for the intermediate content anonymization step: 
\begin{enumerate}[leftmargin=*,itemsep=0pt]
    \item \textbf{Utterance-by-utterance paraphrasing (GPT-4o-mini)}: Each utterance is independently paraphrased.
    \item \textbf{Segment-based paraphrasing (Gemma-3-4B, GPT-5)}: We paraphrase segments of text spanning multiple utterances (using either 16 utterances or roughly 300 tokens) to better capture and alter broader discourse patterns.
\end{enumerate}
For both approaches, we systematically vary the types of prompts used to guide the paraphrasing and experiment with different underlying generative models, from smaller, locally-run models to large, API-based language models. The formulation of these prompts, which includes instructions to condense content and alter utterance length in addition to paraphrasing and PII removal, is a critical factor in successfully mitigating the content-based attack while preserving semantic meaning.\footnote{Our source code and prompts are here: \scriptsize{\url{https://github.com/caggazzotti/long-form-speech-anonymization}}.} To contextualize the paraphrases, we provide a window of utterances from the previous $N$ utterances in the sequence in the prompt ($N=8$ in our experiments).

\begin{figure}[t]
\centering
\includegraphics[width=\columnwidth]{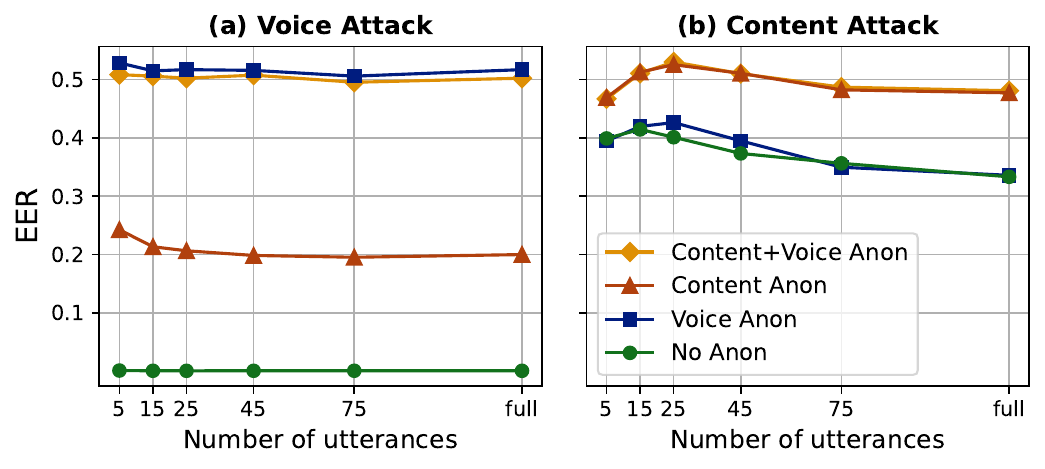} \vspace{-.7cm}
\caption{Left: While voice-only anonymization (using an ASR-TTS pipeline) successfully defeats an audio-only attacker (high EER), it remains vulnerable to a content-only attacker~\cite{aggazzotti2024can}. Right: The content attacker's performance improves as more utterances from the same speaker are aggregated~\cite{aggazzotti2025asr}.}\vspace{-.4cm}
\label{fig:preliminary_results}
\end{figure}

\section{Experiments}
Our experimental framework is designed to evaluate the vulnerability of different anonymization strategies to content-based attacks. The results shown in~\autoref{fig:preliminary_results} demonstrate the core problem: a standard ASR-TTS pipeline, which perfectly anonymizes the acoustic voice, remains highly vulnerable to a content-only attacker. The attacker's performance improves (lower EER) as more utterances from the same speaker are aggregated, confirming that linguistic style can be a biometric identifier~\cite{aggazzotti2024can,aggazzotti2025asr,rivera-soto2021}. Building on this, we conduct a comparative study of three anonymization strategies within the ASR-TTS pipeline:
\begin{enumerate}[leftmargin=*,itemsep=0pt]
    \item \textbf{Audio-only Anonymization}: Only voice anonymization, no content modification. This serves as our primary baseline.
    \item \textbf{Content-only Anonymization}: Our proposed method involves paraphrasing the content to remove PII and alter its style.
    \item \textbf{Voice+Content Anonymization}: Paraphrase the content and then synthesize the paraphrases using an anonymized voice to obtain comprehensive speaker privacy.
\end{enumerate}
This series of experiments allows us to quantify the trade-offs between the level of content modification and the degree of privacy protection, providing a clear picture of the necessity of full stylistic anonymization for long-form audio.

\subsection{Experimental Details}
Our primary corpus is the Fisher Speech Corpus~\cite{cieri2004fisher}, which contains almost 2000 hours of conversational telephone speech. To evaluate the effectiveness of our proposed content anonymization techniques, we adopt the experimental speaker verification setup from recent work in speaker attribution on transcribed speech~\cite{aggazzotti2024can}. This framework is designed to test the robustness of attribution models against topic variation, which is a critical factor in long-form audio. Specifically, we choose their `hard' setting (1944 total trials): positive trials (959) consist of conversations from the same speaker, with each conversation assigned a different conversational topic; negative trials (985) consist of conversations from different speakers but are restricted to the same topic. This design eliminates reliance on topic cues and forces the attribution model to rely on more subtle, stylistic features of speakers’ language. Note that the negative trials are under-sampled compared to an all-pairs trials setup. For the anonymization experiments, the first side of the trial is kept as-is (no anonymization), while the second side has voice-only, content-only, or both voice+content anonymization. 

\vspace{3pt}\noindent \textbf{Voice anonymization} We first employ Whisper-medium\footnote{\scriptsize{\url{https://github.com/openai/whisper}}} as the ASR system, followed by speech synthesis using the multilingual zero-shot TTS model XTTS~\cite{casanova24_interspeech}. (The DTW scores in~\autoref{table:alisim} confirm semantic preservation between the original and anonymized transcripts.) For each speaker in the evaluation set, we generate a pseudo target speaker embedding by sampling 5–6 speakers from the VoxCeleb2~\cite{chung18b_interspeech} corpus and computing a randomly weighted combination of their speaker embeddings. For each of these speakers, we use their longest available utterance to compute the embedding. Note that for speakers participating in multiple conversations, the same pseudo target speaker is consistently assigned to ensure cross-conversation consistency. 

For each trial in the Fisher corpus, only the second speaker's utterances are anonymized using the ASR–TTS pipeline, while the first speaker’s utterances remain non-anonymized. The anonymized audio is subsequently transcribed (still using Whisper-medium) into text and all capitalization and punctuation (other than apostrophes and hyphens) are removed to prevent any signal from transcription style. The content-based attribution attack system is then applied to these normalized transcriptions of the voice-anonymized trials.

\vspace{3pt}\noindent \textbf{Content anonymization} The Whisper transcriptions of the second side of the verification trials are PII-sanitized and paraphrased using each LLM model (in one pass). The content attack operates on trials of a Whisper transcription on one side paired with the now-anonymized side on the other.  To control for content in the voice attack (\autoref{fig:preliminary_results}), we match the source and target speakers during synthesis and use paraphrased transcripts. Speaker verification is then performed between the original and synthesized audio.

\vspace{3pt}\noindent \textbf{Full anonymization} To anonymize both voice and content, the paraphrases are instead synthesized using the pseudo target speaker embedding as described above. These synthesized trials are then re-transcribed for the content attack. 

\vspace{-.1cm}
\subsection{Attack models}
For the voice attack, we utilize the WavLM-Base for speaker verification model\footnote{\scriptsize{\url{https://huggingface.co/microsoft/wavlm-base-sv}}} to extract speaker embeddings. 
For the content attack, we introduce Sentence LUAR (\textsc{SLUAR}\footnote{\scriptsize{\url{https://huggingface.co/noandrews/sluar}}}), which follows the Learning Universal Authorship Representations (\textsc{LUAR}) recipe~\cite{rivera-soto2021}, a top performing authorship attribution model trained on a large Reddit dataset~\cite{khan2021} that employs more invariant stylistic features to identify authors, except we preprocess input documents into sentences to reduce the distribution shift to speech transcript utterances. 

\vspace{-.1cm}
\subsection{Evaluation Metrics}
The evaluation of content anonymization requires a multi-faceted approach. The primary metric for privacy protection is the EER of the content-based attacker under the `hard' setting. A higher EER is desirable as it indicates that the attacker is less able to distinguish between speech from the same speaker and speech from different speakers. An EER of 50\% represents the ideal scenario where the attacker's performance is no better than random chance.

However, EER alone is insufficient. A successful content anonymization system should not only obscure the speaker's identifying stylistic features but also produce content that is fluent, semantically coherent, and not readily identifiable as machine-generated. To this end, we introduce a second evaluation metric: the \textbf{detectability} of the anonymized text. We use a state-of-the-art zero-shot machine-generated text detector, Binoculars~\cite{hans2024spotting}, to assess the extent to which the anonymized content can be distinguished from human-written text. A low detectability score is desirable, as it indicates that the paraphrasing has produced natural-sounding language that does not contain obvious machine-generated artifacts.

We also evaluate the utility of the anonymized speech both for its audio naturalness and its similarity to the original content. Naturalness is assessed using UTMOS~\cite{saeki2022utmos}, a model trained to automatically predict mean opinion scores (MOS) on a 1–5 scale for synthesized or processed speech, where higher scores indicate better perceptual quality.

To ensure the paraphrasing retains the original content, we measure the semantic similarity between the original and paraphrased content using greedy alignment scores and Dynamic Time Warping (DTW) similarity scores. These metrics account for the fact that the number of paraphrased utterances may be different than the number of original utterances by seeking an alignment between them, based on the semantic similarity between all utterances~\cite{reimers-2019-sentence-bert}. For greedy alignment, the final score is the mean similarity of all the identified pairs; this rewards high-quality one-to-one semantic mappings. For DTW, the score is the normalized cumulative negated cost of the optimal alignment.

\vspace{-.2cm}
\subsection{Privacy}
\noindent \textbf{Impact of number of utterances on privacy}
\autoref{fig:preliminary_results} compares the effectiveness of voice and content attacks using different anonymization strategies. The key findings are that (1) the content-based attack improves (lowers EER) with the number of utterances, and (2) content-based anonymization effectively ``flattens'' the curve, rendering the content-based attack significantly less effective as the number of utterances available to the attacker increases. Note that the relatively low EER observed under the content-only anonymization in the voice attack is expected, as the speaker identity remains the same, and a speaker verification model is primarily sensitive to speaker-specific acoustic properties, not textual content. When both the transcript is paraphrased and the target speaker is changed, following the voice+content anonymization procedure, the EER exceeds 50\%.

\begin{figure}[t]
\centering
\includegraphics[width=0.65\columnwidth]{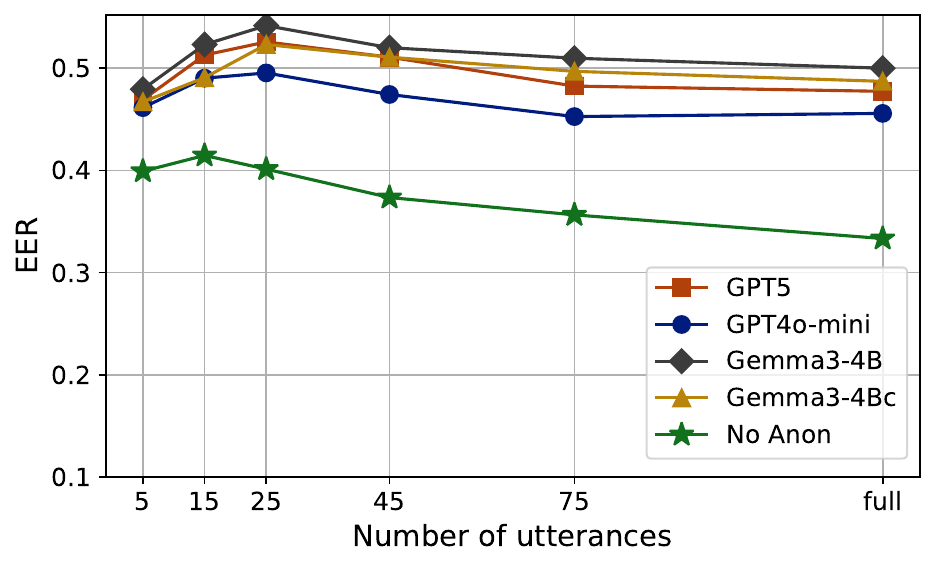}\vspace{-.4cm}
\caption{EER performance of API-based and local LLM paraphrasing models against a content-based attribution attack. All paraphrasers reduce attribution performance to around chance. Segment-based paraphrasing (GPT5, Gemma3s) preserves privacy more than utterance-based paraphrasing (GPT4o-mini).}\vspace{-.4cm}
\label{fig:paramdls}
\end{figure}

\vspace{3pt} \noindent \textbf{Paraphrasing model comparison}
A key component of our proposed method is the generative model used for paraphrasing. We compare two distinct families of models for this task---large, API-based language models (e.g., GPT-5) and smaller, locally-run open-source models (e.g., Gemma-3-4B)---to evaluate the fundamental trade-offs among model capability, privacy, and practicality. 
\begin{itemize}[leftmargin=*,itemsep=0pt]
    \item \textbf{API-based Models (e.g., GPT-5)}: While offering high-quality paraphrasing, they introduce a privacy risk by requiring data to be sent to a third-party service~\cite{openai_2025_gpt5}.
    \item \textbf{Local Models (e.g., Gemma-3-4B)}: These models ensure privacy by keeping data on-device and eliminating API costs, but their ability to sufficiently anonymize content is a key question we investigate~\cite{team2025gemma}. 
\end{itemize}
Since previous work has shown that utterance length can be a useful indicator for content-based attribution models~\cite{aggazzotti2025asr}, we prompt the LLMs to condense the content, specifically changing the utterance length in addition to changing the style. They are also instructed to replace any PII with fictional ones but align with the gender of the original speaker.~\autoref{fig:paramdls} shows the results for both model types against the content-based attack. All paraphrasers make content-based attribution more difficult, especially the segment-based paraphrasers that capture broader speaker style. GPT4o-mini, the only utterance-by-utterance paraphraser, is the least robust. 

\subsection{Utility}
\textbf{Audio Naturalness (UTMOS)}
To evaluate the naturalness of the anonymized audio, we compute UTMOS \cite{saeki2022utmos} scores for both the original and anonymized utterances. Anonymized speech using both audio-only and full voice+content anonymization achieves an UTMOS score of $3.14$, substantially higher than original ground-truth audio samples with an UTMOS score of $2.09$. This difference is likely because the TTS models are trained on high-quality clean speech, whereas the original Fisher recordings consist of spontaneous, conversational audio. While this results in cleaner-sounding speech, the TTS system may also remove conversational disfluencies, which could be important in downstream tasks like attribution.

\vspace{5pt} \noindent \textbf{Content Similarity}
\autoref{table:alisim} reports semantic preservation metrics for the paraphrases. For reference, we also tried smaller (1B) models that failed to follow paraphrasing instructions, which obtained around $0.5$ similarity under these metrics, while two identical transcripts should achieve a perfect score of 1. As shown in the table, the paraphrasing methods tend to reduce the average utterance length compared to the average $9.44$ tokens of the original transcripts. 

\begin{table}[!h]\vspace{-.3cm}
    \footnotesize
    \caption{Greedy alignment scores (GAS), DTW similarity scores (DTW-Sim), and mean utterance length for each LLM paraphraser. }
    \label{table:alisim}\vspace{-.2cm}
    \begin{center}
    \begin{sc}
    \begin{tabular}[c]{@{\ \ \ }l@{\ \ \ }ccc@{\ \ \ }}
        \toprule
        \textbf{LLM paraphraser} & \textbf{GAS} $\uparrow$ & \textbf{DTW-Sim} $\uparrow$ & \textbf{Mean utt. len.}\\
        \midrule
        Gemma3-4B & 0.648 & 0.582 & 7.78\\
        Gemma3-4Bc & 0.647 & 0.637 & 7.58\\
        GPT4o-mini & 0.678 &  0.702 & 9.82 \\
        GPT5 & \textbf{0.699}  & \textbf{0.739} & 5.55 \\ 
        \bottomrule
    \end{tabular}\vspace{-.4cm}
    \end{sc}
    \end{center}
\end{table}

\noindent \textbf{Detectability}
To evaluate how fluent and coherent the anonymized speech and content are, we use both a synthetic speech detector, the state-of-the-art SSL-AASIST \cite{tak22_odyssey}, and a synthetic text detector, the state-of-the-art Binoculars ~\cite{hans2024spotting}. These detectors are zero-shot, operating on individual utterances to produce a score indicating how likely the utterance is AI-generated. For each increasing increment of utterances, we take the mean of the scores for the individual utterances in that set. We then calculate the EER over those scores for the paired real and synthetic trials. 

\begin{figure}[t]
\centering
\includegraphics[width=\columnwidth]{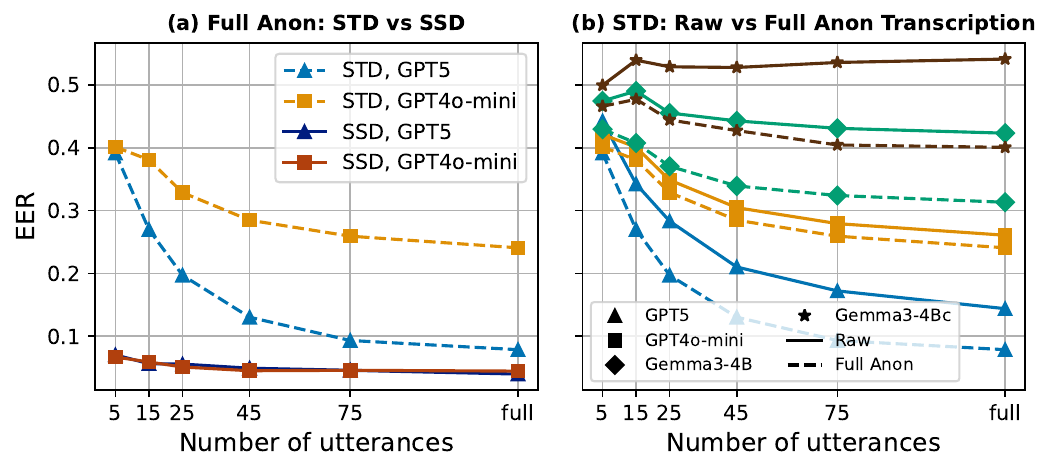} \vspace{-.8cm}
\caption{Left: Synthetic speech detection (SSD) is significantly better than synthetic text detection (STD) on fully anonymized speech/text, especially with only a few utterances. Right: Better STD on original (raw) paraphrases suggests that synthesis/re-transcription removes some machine style.}\vspace{-.4cm}
\label{fig:detRes}
\end{figure}

The left graph in \autoref{fig:detRes} compares the synthetic speech detector (SSD) to the synthetic text detector (STD) for the two API-based models. The performance of the SSD is comparatively better than that of the STD and is highly accurate even with only a few utterances. The text detector is not successful with fewer utterances, but improves significantly with more text. The right graph compares text detection on the raw paraphrases (original Whisper transcriptions) versus the fully anonymized transcriptions (paraphrased, synthesized with a random voice, and re-transcribed). The raw paraphrases are easier to detect as machine-generated, suggesting that synthesizing and re-transcribing removes some machine-generated style. Also, the Gemma3 models are significantly more difficult to detect, especially and unsurprisingly the conservative model (Gemma3-4Bc), which keeps half of the utterances the same as the original. \vspace{-.2cm}

\section{Conclusion}
Existing voice anonymization techniques are built on the flawed premise that obscuring a speaker's voice is sufficient to protect their identity. This assumption breaks down in real-world scenarios involving long-form audio, where the linguistic content can be exploited by an attacker~\cite{aggazzotti2024can}. Our work highlights this ``content gap'' compared to voice-only anonymization and attribution and shows how to close it by incorporating a content anonymization step in an ASR-TTS pipeline.

\vspace{-.2cm}\subsection{Limitations and Future Work}
The proposed approach has several limitations that suggest avenues for future research. First, the reliance on an ASR-TTS pipeline, though effective for anonymization, introduces potential challenges. Errors in the initial ASR stage can propagate through the pipeline, potentially affecting the quality and fluency of the final synthesized speech~\cite{Horiguchi2022EncoderDecoderBA}. Future work could explore end-to-end models that perform joint voice+content anonymization in a single step, potentially mitigating these issues. Second, the paraphrasing component, while essential for anonymizing the linguistic content, may not perfectly preserve the original semantic meaning in all cases. Subtle nuances, sarcasm, or specific turns of phrase may be altered or lost in the paraphrasing process. Further research is needed to develop more robust paraphrasing models that can better navigate the trade-off between stylistic anonymization and semantic fidelity. Finally, content-based attacks may prove even more effective in semi-informed attacker settings where anonymized data is available to improve the attack model, which may in turn require different anonymization strategies.

\section{Acknowledgments} \vspace{-.2cm}
This research is supported in part by the Office of the Director of National Intelligence (ODNI), Intelligence Advanced Research Projects Activity (IARPA), via the HIATUS Program contract \#D2022-2205150003 and the ARTS Program contract \#D2023-2308110001. The views and conclusions contained herein are those of the authors and should not be interpreted as necessarily representing the official policies, either expressed or implied, of ODNI, IARPA, or the U.S. Government. The U.S. Government is authorized to reproduce and distribute reprints for governmental purposes notwithstanding any copyright annotation therein.

\bibliographystyle{IEEEtran}
\bibliography{papers}

@article{aggazzotti2025asr,
    author = {Aggazzotti, Cristina and Wiesner, Matthew and Smith, Elizabeth Allyn and Andrews, Nicholas},
    title = {{The Impact of Automatic Speech Transcription on Speaker Attribution}},
    journal = {Trans. of the Association for Computational Linguistics},
    volume = {13},
    pages = {1578-1596},
    year = {2025},
}

@misc{openai_2025_gpt5,
author = {OpenAI},
title = {{GPT-5 System Card}},
year = {2025},
month = {August},
}

@inproceedings{reimers-2019-sentence-bert,
  title = {{Sentence-BERT: Sentence Embeddings using Siamese BERT-Networks}},
  author = "Reimers, Nils and Gurevych, Iryna",
  booktitle = "Proc. of EMNLP",
  year = "2019",
}

@article{team2025gemma,
  title={{Gemma 3 Technical Report}},
  author={Kamath, Aishwarya and Ferret, Johan and Pathak, Shreya and Vieillard, Nino and Merhej, Ramona and Perrin, Sarah and Matejovicova, Tatiana and others},
  journal={arXiv:2503.19786},
  year={2025}
}

@inproceedings{williams2021revisiting,
  title={{Revisiting Speech Content Privacy}},
  author={Williams, Jennifer and Yamagishi, Junichi and No{\'e}, Paul-Gauthier and Botinhao, Cassia Valentini and Bonastre, Jean-Fran{\c{c}}ois},
  booktitle={Proc. of ISCA-SPSC},
  year={2021},
}

@inproceedings{williams2023new,
  title={{New Challenges for Content Privacy in Speech and Audio}},
  author={Williams, Jennifer and Pizzi, Karla and Das, Shuvayanti and No{\'e}, Paul-Gauthier},
  booktitle={Proc. of ISCA-SPSC},
  year={2023},
}

@inproceedings{williams2024,
  author={Williams, Jennifer and Pizzi, Karla and Noe, Paul-Gauthier and Das, Sneha},
  booktitle={Speech Communication; 15th ITG Conf.}, 
  title={{Exploratory Evaluation of Speech Content Masking}}, 
  year={2023},
  volume={},
  number={},
  pages={215-219},}

@inproceedings{turan2022,
    title = {{Adapting Language Models When Training on Privacy-Transformed Data}},
    author = "Turan, Tugtekin  and
      Klakow, Dietrich  and
      Vincent, Emmanuel  and
      Jouvet, Denis",
    booktitle = "Proc. of the 13th Language Resources and Evaluation Conference",
    year = "2022",
    pages = "4367--4373"
}

@inproceedings{rivera-soto2021,
    title = {{Learning Universal Authorship Representations}},
    author = "Rivera-Soto, Rafael A.  and
      Miano, Olivia Elizabeth  and
      Ordonez, Juanita  and
      Chen, Barry Y.  and
      Khan, Aleem  and
      Bishop, Marcus  and
      Andrews, Nicholas",
    booktitle = "Proc. of EMNLP",
    year = "2021",
    pages = "913--919",
}

@article{tomashenko2024voiceprivacy,
  title={{The VoicePrivacy 2024 Challenge Evaluation Plan}},
  author={Tomashenko, Natalia and Miao, Xiaoxiao and Champion, Pierre and Meyer, Sarina and Wang, Xin and Vincent, Emmanuel and Panariello, Michele and Evans, Nicholas and Yamagishi, Junichi and Todisco, Massimiliano},
  journal={arXiv preprint arXiv:2404.02677},
  year={2024}
}

@inproceedings{hui2025securespeech,
  title={{{SecureSpeech}: Prompt-based Speaker and Content Protection}},
  author={Hui, Belinda Soh Hui and Miao, Xiaoxiao and Wang, Xin},
  booktitle ={IEEE International Joint Conference on Biometrics},
  year={2025}
}

@inproceedings{li2024best,
    author = {Xinyuan, Henry Li and Cai, Zexin and Garg, Ashi and Garcia-Perera, Leibny Paola and Duh, Kevin and Khudanpur, Sanjeev and Andrews, Nicholas and Wiesner, Matthew},
    booktitle = {Proc. of ISCA-SPSC},
    title = {{{HLTCOE} Submission to the 2024 {VoicePrivacy Challenge}}},
    year = {2024}
}

@article{aggazzotti2024can,
    author = {Aggazzotti, Cristina and Andrews, Nicholas and Smith, Elizabeth Allyn},
    journal = {Trans. of the Association for Computational Linguistics},
    pages = {875--891},
    title = {{Can Authorship Attribution Models Distinguish Speakers in Speech Transcripts?}},
    volume = {12},
    year = {2024}
}

@INPROCEEDINGS{li2025attack,
  author={Xinyuan, Henry Li and Garg, Ashi and Cai, Zexin and Duh, Kevin and Garc{\'i}a-Perera, Leibny Paola and Khudanpur, Sanjeev and Andrews, Nicholas and Wiesner, Matthew},
  booktitle={Proc. of IEEE ICASSP},
  title={{{HLTCOE} Submission to the {VoicePrivacy Attacker Challenge}}},
  year={2025},
  pages={1-2},
}

@inproceedings{meyer2025use,
   title={{Use Cases for Voice Anonymization}},
   author={Meyer, Sarina and Vu, Ngoc Thang},
   booktitle={Proc. of ISCA-SPSC},
   year={2025},
}

@inproceedings{saeki2022utmos,
  title={{UTMOS: UTokyo-SaruLab System for VoiceMOS Challenge 2022}},
  author={Saeki, Tomoki and Kawamura, Wataru and Takamichi, Shinnosuke and Saruwatari, Hiroshi},
  booktitle={Proc. of the VoiceMOS Challenge 2022},
  pages={33--38},
  year={2022}
}

@article{Horiguchi2022EncoderDecoderBA,
  title={{Encoder-Decoder Based Attractors for End-to-End Neural Diarization}},
  author={Shota Horiguchi and Yusuke Fujita and Shinji Watanabe and Yawen Xue and Paola Garc{\'i}a},
  journal={IEEE/ACM Trans. on Audio, Speech, and Lang. Processing},
  year={2022},
  volume={30},
  pages={1493-1507}
}

@inproceedings{cieri2004fisher,
  title={{The {F}isher {C}orpus: A Resource for the Next Generations of Speech-to-Text}},
  author={Cieri, Christopher and Miller, David and Walker, Karen},
  booktitle={Proc. of the 4th International Conference on Language Resources and Evaluation},
  pages={69--71},
  year={2004}
}

@inproceedings{hans2024spotting,
author = {Hans, Abhimanyu and Schwarzschild, Avi and Cherepanova, Valeriia and Kazemi, Hamid and Saha, Aniruddha and Goldblum, Micah and Geiping, Jonas and Goldstein, Tom},
title = {{Spotting {LLMs} with {B}inoculars: Zero-shot Detection of Machine-generated Text}},
year = {2024},
booktitle = {Proc. of ICML},
articleno = {698},
numpages = {19},
}

@inproceedings{casanova24_interspeech,
  title = {{XTTS: a Massively Multilingual Zero-Shot Text-to-Speech Model}},
  author = {Edresson Casanova and Kelly Davis and Eren Gölge and Görkem Göknar and Iulian Gulea and Logan Hart and Aya Aljafari and Joshua Meyer and Reuben Morais and Samuel Olayemi and Julian Weber},
  booktitle = {{Interspeech 2024}},
  pages = {4978--4982},
}

@inproceedings{chung18b_interspeech,
  title     = {{{VoxCeleb2}: Deep Speaker Recognition}},
  author    = {Joon Son Chung and Arsha Nagrani and Andrew Zisserman},
  booktitle = {Interspeech 2018},
  pages     = {1086--1090},
}

@inproceedings{tak22_odyssey,
  title     = {{Automatic Speaker Verification Spoofing and Deepfake Detection Using Wav2vec 2.0 and Data Augmentation}},
  author    = {Hemlata Tak and Massimiliano Todisco and Xin Wang and Jee-weon Jung and Junichi Yamagishi and Nicholas Evans},
  year      = {2022},
  booktitle = {Speaker and Language Recognition Workshop (Odyssey)},
  pages     = {112--119},
}

@article{stamatatos2009,
  title={{A Survey of Modern Authorship Attribution Methods}},
  author={Stamatatos, Efstathios},
  journal={Journal of the Amer. Society for Information Science and Technology},
  volume={60},
  number={3},
  pages={538--556},
  year={2009},
}

@article{neal2017,
  title={{Surveying Stylometry Techniques and Applications}},
  author={Neal, Tempestt and Sundararajan, Kalaivani and Fatima, Aneez and Yan, Yiming and Xiang, Yingfei and Woodard, Damon},
  journal={ACM Computing Surveys},
  volume={50},
  number={6},
  pages={1--36},
  year={2017},
}

@inproceedings{wegmann2022,
    title = {{Same Author or Just Same Topic? {T}owards Content-Independent Style Representations}},
    author = "Wegmann, Anna  and
      Schraagen, Marijn  and
      Nguyen, Dong",
    booktitle = "Proc. of the 7th Workshop on Representation Learning for NLP",
    year = "2022",
    pages = "249--268",
}

@inproceedings{khan2021,
    title = {{A Deep Metric Learning Approach to Account Linking}},
    author = "Khan, Aleem  and
      Fleming, Elizabeth  and
      Schofield, Noah  and
      Bishop, Marcus  and
      Andrews, Nicholas",
    booktitle = "Proc. of NAACL: Human Lang. Technologies",
    year = "2021",
    pages = "5275--5287"
}

@inproceedings{baocarpuat2024,
    author = {Calvin Bao and Marine Carpuat},
    title = {{{Keep it Private: U}nsupervised Privatization of Online Text}},
    booktitle = {Proc. of NAACL: Human Lang. Technologies (Vol 1: Long Papers)},
    year = 2024,
    pages = {8678–8693},
}

@inproceedings{fisher2024jamdec,
    title = "{{JAMDEC}: Unsupervised Authorship Obfuscation using Constrained Decoding over Small Language Models}",
    author = "Fisher, Jillian  and
      Lu, Ximing  and
      Jung, Jaehun  and
      Jiang, Liwei  and
      Harchaoui, Zaid  and
      Choi, Yejin",
    booktitle = "Proc. of NAACL: Human Lang. Technologies (Vol 1: Long Papers)",
    year = "2024",
    pages = "1552--1581",
}

\end{document}